\documentclass[aps,prd,showpacs,amssymb,eqsecnum,amsfonts,epsf,
preprintnumbers,nofootinbib,superscriptaddress,twocolumn]{revtex4}
\usepackage[dvips]{graphicx}
\usepackage{bm,latexsym,amsmath,amssymb,amsfonts}
\usepackage{graphicx}
\usepackage{natbib}

\newcommand*{\cH}{{\cal H}}

\newcommand{\beq}{\begin{equation}}
\newcommand{\eeq}{\end{equation}}
\newcommand*{\tb}{{\tilde B}}

\begin{document}

\title{Cosmological evolution of regularized branes in 6D warped flux compactifications}

\author{Masato~Minamitsuji}
\email[Email: ]{masato"at"theorie.physik.uni-muenchen.de}
\affiliation{Arnold-Sommerfeld-Center for Theoretical Physics, Department f\"{u}r Physik, Ludwig-Maximilians-Universit\"{a}t, Theresienstr. 37, D-80333, Munich, Germany}
\author{David~Langlois}
\affiliation{
Laboratoire APC, B$\hat{a}$timent Condorcet, 10,
Rue Alice Domon et L\'eonie Duquet,
75205 Paris Cedex 13}
\affiliation{Institut d'Astrophysique de Paris, GReCO, CNRS, 98bis boulevard Arago, 75014 Paris, France.}

\begin{abstract}
We study the cosmological evolution of extended branes in 6D warped flux compactification models. 
The branes are endowed with the three ordinary spatial dimensions, which are assumed to be homogeneous and isotropic, as well as an internal extra dimension compactified on a circle.
We embed these codimension 1 branes in a static bulk 6D spacetime, 
whose geometry is a solution of 6D Einstein-Maxwell or Einstein-Maxwell-dilaton theories, corresponding
to a warped flux compactification.
The brane matter consists of a complex scalar field which is coupled to the bulk $U(1)$ gauge field.
In both models, we show that there is critical point which
the brane cannot cross as it moves in the bulk.
We study the cosmological behaviour, especially when the brane approaches this critical point or one of the two conical singularities.
In the present setup where the bulk geometry is fixed, we find that the brane cosmology does not coincide 
with the standard one in the low energy limit.
\end{abstract}

\preprint{LMU-ASC 47/07}
\date{\today}
\maketitle


\section{Introduction}

Braneworld models have been studied actively in the last decade (see e.g. \cite{reviews} for a few reviews on this topic). 
Among models which take explicitly into account
the self-gravity of the branes, i.e. the backreaction on the geometry due to the presence of the brane, 
most efforts have been devoted to the study of gravity and cosmology in
 five dimensional (codimension 1) braneworld models, before (see \cite{BDL}) and especially after the stimulating proposals by Randall and Sundrum
\cite{RS}. However,
if one is interested in a braneworld model in a six or higher dimensional bulk spacetime, one needs to consider a brane with codimension higher than 1.

Among higher codimensional braneworld models,
some significant attention has been paid to codimension 2 branes
(see e.g. Ref. \cite{Papa2} and references therein), because
they possess the  property that, even with some non zero vacuum energy, 
 the brane geometry can be flat, while the bulk geometry surrounding the brane is characterized by a deficit angle. Codimension 2 branes have thus been studied in the context of the cosmological constant problem \cite{cc,cc2,Vinet:2005dg}, but they are also interesting on their own as the simplest case beyond
 codimension 1 branes.

One difficulty however with codimension 2 brane is that the energy momentum tensor is in general restricted to be  that of a pure tension.
To consider more general matter, such as would be relevant to study cosmology in the brane, one needs to regularize the brane. One well-motivated way of regularization is to replace the codimension 2 brane by a thin ring-like codimension 1 brane wrapped around the symmetry axis, as
was firstly suggested in \cite{PST} in the context of the
"rugby-ball"  braneworld model.
The rugby-ball model has a 2D
compact bulk compactified by a magnetic flux
and codimension two branes are located at poles.
This model was originally introduced in order to 
resolve the cosmological constant problem~\cite{cc}, but it was also
pointed out that the so-called selftuning property did not actually
work \cite{cc2,Vinet:2005dg} (see also \cite{Koyama:2007rx} for a review).

In the present work, we consider static warped 6D braneworld solutions \cite{Mukohyama, 6d_sugra} with a 
 $U(1)$ gauge field coupled to gravity,
which are generalizations of the rugby-ball model.
Only static regularized branes have been
considered in these geometries so far as now \cite{PPZ, KM, BHT}.
The goal of the present work is to study the cosmological behavior of branes
in such 6D braneworld models, i.e.
to consider the motion of the branes in these given bulk geometries.
\footnote{In Ref. \cite{Vinet:2005dg}, cosmology in such models has been discussed by employing a different way of regularization.}
We basically follow the well-established method to discuss cosmology in
 5D braneworld models, using   the static bulk point of view \cite{CR, Kraus, KK, Ida} rather than 
the brane-based approach \cite{BDL}.
In order to satisfy  all the junction conditions, we assume, like in \cite{PST} that 
the branes contain a complex scalar field
but we allow for a time evolution of its radial component.

The paper is organized as follows. 
In Sec. II, we give a general formalism to discuss cosmology on the brane.
In Sec. III, we focus on the cosmological evolution of the brane in 
the models based on the 6D Einstein-Maxwell theory \cite{Mukohyama}.
In Sec. IV,
we discuss the case of the 6D Einstein-Maxwell-dilaton model
(more precisely the bosonic part of Nishino-Sezgin supergravity \cite{N-S})
\cite{6d_sugra} as the same manner as the line of the previous section.
In Sec. V, we give summary and conclusions before closing the article.

\section{Basic formulation}

Before considering explicit  models, we study in this section the general 
problem of the motion of a codimension 1 brane embedded in a 6D static geometry with axial symmetry in its two internal dimensions. 
We start from a static 6D metric of the form
\begin{eqnarray}
ds^2=A^2(w) (-dt^2+d{\bf x}^2)
     +B^2(w)dw^2+C^2(w)d\theta^2\,.\label{metric_ge}
\end{eqnarray}
Here $\theta$ is implicitly assumed to be an angular coordinate, 
associated with a compactified internal dimension. 
$A(w)$ is assumed to be a monotonically increasing function of $w$.
We embed into this static bulk a codimension 1 brane,  which we assume to be homogeneous and isotropic along 
the three ordinary spatial dimensions as well as along the angular coordinate $\theta$. The position of the 
brane is thus simply characterized by its coordinates $t(\tau)$ and $w(\tau)$, 
given for example as functions of the proper time $\tau$. By definition of the proper time, one has the following 
relation
\begin{eqnarray}
-A^2 \dot{t}{}^2+B^2\dot{w}{}^2=-1\,.
\label{normalization}
\end{eqnarray}
where a dot denotes a derivative with respect to $\tau$. 
The above relation comes from the normalization 
of the brane 5-velocity, whose components are given by 
\begin{eqnarray}
&&
u^{t}=\dot{t}\,\quad 
u^{w}=\dot{w}\,\quad
u_{t}=-A^2\dot{t}\,\quad
u_{w}=B^2\dot{w}\,.
\end{eqnarray}
The unit normal vector to the brane is defined by
\begin{eqnarray}
&&
n^{w}=\frac{A}{B}\dot{t}\,\quad
n^{t}=\frac{B}{A}\dot{w}\,
\nonumber\\
&&n_{w}=AB\dot{t}\quad
n_{t}=-AB \dot{w}.
\end{eqnarray}
If $q_{ab}$ denotes the induced metric on the brane,
the extrinsic curvature tensor is defined by 
\beq
 K_{ab}=q_{a}^{\ c}\nabla_cn_b=\frac{1}{2}{\cal L}_{\bf n} q_{ab},
\eeq
where ${\cal L}_{\bf n}$ is the Lie derivative with respect to the normal vector $n^a$.
In order to write explicitly the junction conditions, it is more convenient to  introduce the combination
\beq
\hat K_{ab}=K_{ab}-K q_{ab},
\eeq
where $K=q^{ab}K_{ab}$. The junction conditions are then given by
\beq
\left[ \hat K_{ab}\right]=-\frac{1}{M_6^4} T_{ab}
\eeq
where $T_{ab}$ is the energy momentum-tensor confined on the brane and
$M_6$ is the 6D Planck mass.
We shall set $M_6=1$  unless stated otherwise explicitly.
The brackets on the left hand side denote the jump across the brane of the extrinsic 
curvature tensor, that is the difference between its values on the two sides of the brane.

In our case, since we assume homogeneity of the brane matter as well as isotropy in the three ordinary spatial dimensions, the brane energy momentum tensor is necessarily of the form:
\begin{eqnarray}
  T_{ab}
= \Sigma\, u_a u_b
 +p\,  h_{ab}
 +p_{\theta}\, \theta_a\theta_b,
 \label{EMT}
\end{eqnarray}
where $h_{ab}$ is the three-dimensional metric in the ordinary spatial directions and  $\theta_a$ is 
the unit vector along the internal $\theta$ direction.
$\Sigma$ is the energy density and $p$ and $p_\theta$ are 
respectively the ordinary and internal pressures.

One consequence of the brane junction conditions is  the generalized  energy conservation law on the brane, which can be expressed as
\begin{eqnarray}
 && \dot{\Sigma}
+3\frac{\dot{A}}{A}\big(\Sigma+p\big)
+ \frac{\dot{C}}{C}\big(\Sigma+p_{\theta}\big)
=\big[{\cal T}_{ab}n^a u_b\big]\,,
\nonumber\\
&& \big[{\cal T}_{ab}n^a u_b\big]
=  \big({\cal T}_{ab}n^a u_b\big)_{w=w_{(b)}^+}
  -\big({\cal T}_{ab}n^a u_b\big)_{w=w_{(b)}^-}
\end{eqnarray}
where  ${\cal} T_{AB}$ denotes the bulk energy-momentum tensor, and all quantities are
evaluated at the brane location. The above expression differs from the
usual cosmological conservation law in two respects:  first,  there is  an additional term due the evolution of the internal spatial direction; second, one finds in general on the right hand side a brane-bulk energy exchange term, similar to what can been found
in  5D brane cosmology (see e.g.  \cite{Langlois_Sorbo}).


\section{Einstein-Maxwell model}

\subsection{Bulk spacetime in 6D Einstein-Maxwell theory}

We now discuss the dynamics of an extended brane in the 6D 
Einstein-Maxwell theory. The action in the bulk is given by 
\beq
S_{B}=\int d^6 x\sqrt{-g}
\Big(\frac{1}{2}R
      -\frac{1}{4}F_{AB}F^{AB}
     -\Lambda_0
\Big)\,,
\eeq
where $g_{AB}$ is the 6D metric,
$F_{AB}=2\partial_{[A} A_{B]}$ is the field strength
associated to the $U(1)$ gauge vector $A_B$,
and $\Lambda_0$ is a cosmological constant.

By a double Wick rotation of the 6D Reissner-Nordstroem solution,
one can obtain the solution \cite{Mukohyama}:
\begin{eqnarray}
ds_{6}^2&= &\rho^2 \eta_{\mu\nu}dx^{\mu}dx^{\nu}
  +\frac{d\rho^2}{F(\rho)}
  +c_0^2 F(\rho) d\varphi^2\,,
\nonumber\\
&& F_{\rho\varphi}=-\frac{b_0}{\rho^4}
\end{eqnarray} 
where
\begin{eqnarray}
F(\rho)
=-\frac{\Lambda_0}{10}\rho^2
 -\frac{b_0^2}{12c_0^2\rho^6}
 +\frac{\mu_0}{\rho^3}\,.
\end{eqnarray}
Assuming that $F$ has two positive roots $\rho_-$ and $\rho_+$ such that 
$0<\rho_-<\rho_+$, one can replace the two parameters $b_0$ and $\mu_0$ by $\rho_+$ and $\alpha\equiv \rho_- /\rho_+$.
Using $F(\rho_\pm)=0$,  $b_0$ and $\mu_0$ are given by
\begin{eqnarray}
b_0^2=\frac{6}{5}
      c_0^2\Lambda_0
       \rho_+^8 \alpha^3
       \frac{1-\alpha^5}{1-\alpha^3}
\,,\quad
\mu_0=\frac{\Lambda_0}{10}\rho_+^5
      \frac{1-\alpha^8}{1-\alpha^3}\,.
\end{eqnarray}
It is also convenient to replace the  coordinates $\rho$ and $\varphi$ by the new coordinates 
$w$ and $\theta$, defined respectively by 
\begin{eqnarray}
\rho=\rho_+ w
\,,\quad
\varphi=\frac{1}{\Lambda_0 \rho_+ (1-\alpha)}\theta\,,
\label{new_coord}
\end{eqnarray}
so that
the bulk
metric  now reads
\begin{eqnarray}
 ds_{6}^2
&=&R_0^2 \Big[
\frac{4dw^2}{(1-\alpha)^2f(w)}
+c_0^2 f(w)d\theta^2
       \Big]
\nonumber\\
&+&\rho_+^2 w^2 \eta_{\mu\nu}dx^{\mu}dx^{\nu}\,,
\label{metric_em}
\end{eqnarray}
with
\begin{eqnarray}
&&f(w)=\frac{1}{5(1-\alpha)^2}
     \Big[
      -w^2 
      -\frac{1}{w^6}
       \frac{\alpha^3(1-\alpha^5)}{1-\alpha^3}
      +\frac{1}{w^3} 
       \frac{1-\alpha^8}{1-\alpha^3}
     \Big]\,,
\nonumber\\
&&
\end{eqnarray}
and the field strength is now given by
\beq
 F_{w\theta}=-\frac{2c_0R_0 S}{(1-\alpha)w^4}, \quad S=S(\alpha)\equiv\sqrt{\frac{3\alpha^3(1-\alpha^5)}{5(1-\alpha^3)}}\,,
\eeq
with $R_0^2\equiv (2\Lambda_0)^{-1}$.

The singularities $\rho=\rho_-$ and $\rho=\rho_+$ now correspond to
 $w=\alpha$ and $w=1$ in the new coordinate system.
By expanding the above metric around
these two singularities, one finds that they are in general 
conical singularities with some corresponding deficit  angles.
These conical singularities can 
be related to the presence of two codimension-2 branes. As first suggested in \cite{PST}, one way 
to regularize these codimension 2 branes is to replace them by two codimension 1 branes, 
with one spatial dimension compactified on a circle at the respective positions $w_->\alpha$ and $w_+<1$. These
codimension 1 branes are the boundaries of regular spherical caps, which end smoothly the spacetime.
The spherical caps can be described by solutions similar to that of the 
main bulk, i.e. 
\begin{eqnarray}
ds_{6,(I)}^2
&=&R_I^2 \Big[
\frac{4dw^2}{(1-\alpha)^2f(w)}
+c_I^2 f(w)d\theta^2
       \Big]
\nonumber \\
&+&\rho_+^2 w^2 \eta_{\mu\nu}dx^{\mu}dx^{\nu}\,,
\nonumber\\
F^{(I)}_{w\theta}&=&-\frac{2c_I R_I S}{(1-\alpha)w^4}
\end{eqnarray}
where the index $I$ takes the values  $I=+$ and $I=-$ to denote respectively the northern cap, $w_+<w<1$,
and the southern cap, $\alpha<w<w_-$.
Note that each bulk region is endowed with different cosmological constants
$\Lambda_{\pm}$. 
The parameters $c_+$ and $c_-$ are chosen so that the geometry at the two poles $w=1$ and $w=\alpha$ is now 
smooth, i.e. with no deficit angle. This implies \cite{PPZ, KM}
\begin{eqnarray}
&&c_+=\frac{20(1-\alpha)(1-\alpha^3)}{5-8\alpha^3+3\alpha^8},\nonumber\\
&&c_-=\frac{20(1-\alpha)(1-\alpha^3)\alpha^4}{5\alpha^8-8\alpha^5+3}.
\label{c_I}
\end{eqnarray}
Moreover, the continuity of the $(\theta,\theta)$ component requires
\begin{eqnarray}
\ell:=R_0 c_0=R_+ c_+=R_-c_-\,.
\label{l}
\end{eqnarray}
For a schematic picture of the bulk configuration,
see e.g., Fig. 1 of Ref. \cite{KM}.

The gauge field in the bulk is solved as
\begin{eqnarray}
&&A^{\rm (N)}_{\theta}:=
\frac{2S\ell}{3(1-\alpha)}
 \Big(\frac{1}{w^3}-1\Big)\,,
\nonumber\\
&&A^{\rm (S)}_{\theta}:=
\frac{2S\ell}{3(1-\alpha)}
 \Big(\frac{1}{w^3}-\frac{1}{\alpha^3}\Big)\,,
\label{vectorpotential}
\end{eqnarray}
expression valid respectively in the northern and southern regions.
Note that in the region of the ``$0$"-bulk
the bulk gauge fields are doubly defined.  This is consistent 
only if the difference is simply a pure gauge term, which implies
the Dirac quantization condition
\begin{eqnarray}
N=\frac{2eS \ell (1-\alpha^3)}
       {3(1-\alpha)\alpha^3}\,,\quad
N=0,\pm 1,\pm 2, \cdots.\label{dirac}
\end{eqnarray}

\subsection{Junction conditions}

The matter content of the brane must be compatible with the bulk solution, i.e. must satisfy the junction 
conditions for the metric and the gauge field. In our case, we consider a complex scalar field  coupled to 
the bulk gauge field with the following action
\begin{eqnarray}
S_{b,\pm}&=&
-\int d^5 x
\sqrt{-q}
\Big(
V(|\phi_{\pm}|)
+\frac{1}{2}
 D_{a}\phi_{\pm} \big(D^a \phi_{\pm} \big)^{\ast}
\Big)
\nonumber\\
&=&-\int d^5 x \sqrt{-q}
\Big(
V(\Phi_{\pm})
+\frac{1}{2}\partial^a\Phi_{\pm}
            \partial_a\Phi_{\pm}
\nonumber\\
&+&\frac{1}{2}
 \Phi_{\pm}^2 
 \big(
\partial^a\sigma_{\pm}-e A^{a}
 \big)
 \big(
\partial_a\sigma_{\pm}-e A_{a}
 \big)
\Big)\,,
\end{eqnarray}
 where $\phi_{\pm}= \Phi_{\pm}(\tau) e^{i\sigma_{\pm}(\theta)}$.
This is a natural 
extension of the brane action adopted in \cite{PST, PPZ, KM}
for static branes, 
with the difference that $\Phi_{\pm}$, which  was frozen at a fixed value in the static case, must now 
be promoted to a time-dependent field.
Hereafter, we will often omit the brane subscripts $\pm$  for brevity. 
The above action implies that the energy density, the external and internal pressures, defined earlier
 in (\ref{EMT}), are given 
explicitly by the following expressions:
\begin{eqnarray}
\label{Sigma}
\Sigma&=&V(\Phi
           )
  +\frac{1}{2}\Phi^2
               C^{-2}
    (\partial_{\theta}\sigma
    -e A_{\theta})^2
  +\frac{1}{2}\dot{\Phi}^2
\, ,\\
\label{p_ext}
p&=& -V(\Phi)
  -\frac{1}{2}\Phi^2
        C^{-2}
    (\partial_{\theta}\sigma
     -e A_{\theta})^2
  +\frac{1}{2}\dot{\Phi}{}^2
\, ,\\
\label{p_int}
p_{\theta} &=& -V(\Phi)
  +\frac{1}{2}\Phi^2 
            C^{-2}
   (\partial_{\theta}\sigma
   -e A_{\theta})^2
   +\frac{1}{2}\dot{\Phi}^2.
\end{eqnarray}
To discuss the dynamics of the $(\pm)$-branes,
we can now apply the general formalism
presented in the previous section.
Comparing the metric (\ref{metric_em})
with  Eq. (\ref{metric_ge}), it is easy to identify
\begin{eqnarray}
&&A(w)^2=\rho_+^2w^2,\quad 
B(w)^2=\frac{4R_I^2}{(1-\alpha)^2f(w)},\,
\nonumber \\ 
&&
C(w)^2= \ell^2 f(w).
\end{eqnarray}
Note that the warp factor $A$ is a monotonically increasing function with
respect to $w$ as  assumed in the previous section. 
Since we are interested in the cosmological evolution of the branes, it is convenient to reexpress the various
bulk components as functions of the brane scale factor 
\beq
a(\tau)\equiv A(w_b(\tau)),
\eeq
instead of the radial coordinate $w$.  In the following, we will use  the function $C(a)$ defined  
by substituting in $C(w)$ the expression of $w$ as a function of $a$, as well as
\begin{eqnarray}
\tilde B(a)
=\left(\frac{dA}{dw}\right)^{-1}B
=\frac{2}{(1-\alpha)\rho_+}\frac{R_I}{\sqrt{f}},
\end{eqnarray}
where the right hand side is evaluated at the brane position.

 Defining the jump across the branes as
$[T]_{\pm}:=\pm (T_{\pm}-T_{0})$, the 
junction conditions for the metric
$[\hat K_{ab}]=-T_{ab}$
are given by 
\begin{eqnarray}
&&\Big(\frac{3}{a}+\frac{C'}{C}\Big)
  \Big[X\Big]
+\Big(\ddot{a}+\frac{\tilde B '}{\tilde B} \dot{a}^2\Big)
  \Big[\frac{1}{X}\Big]
  =p\,,
\label{junction1}\\
&&\frac{4}{a}\Big[X\Big]
+\Big(\ddot{a}+\frac{\tilde B '}{\tilde B} \dot{a}^2\Big)
  \Big[\frac{1}{X}\Big]
  =p_{\theta}
\label{junction2}\\
&&\Big(\frac{3}{a}+\frac{C'}{C}\Big)
  \Big[X\Big]
  =-\Sigma
\,,\label{junction3}
\end{eqnarray}
where $X:=\sqrt{\tilde B{}^{-2}+\dot{a}^2}$
and the prime $'$ denotes the derivative with respect to $a$.
These relations generalize the expressions given in \cite{PST} for a static brane. Note that in the static 
case, the left hand sides of (\ref{junction1}) and (\ref{junction3}) are identical, which implies $p=-\Sigma$.
Note also that in the cosmological case,  the topological condition  is 
the same as in the static case (see Appendix A-1 for details).

The last junction condition (\ref{junction3}) implies  the generalized Friedmann equation
\begin{eqnarray}
&&\frac{\dot{a}^2}{a^2}
=\frac{{\tilde \rho}^2}
{16\pi^2\ell^2 f \big(3+\frac{f'}{2f}a\big)^2} 
\nonumber\\
&&+\frac{\pi^2\ell^2 f}{a^4}
    \frac{\big(3+\frac{f'}{2f}a\big)^2 (\tilde B{}_{\pm}^{-2}-\tilde B{}_0^{-2})^2}
          {{\tilde \rho}^2}
-\frac{\tilde B_{\pm}^{-2}+\tilde B_0^{-2}}{2a^2},
\label{Friedmann}
\nonumber\\
&&
\end{eqnarray}
where $\tilde \rho:= 2\pi\ell\sqrt{f}\Sigma$, obtained by integrating $\Sigma$ over the internal 
dimension,  is the 4D effective energy density.
 On the right hand side, one finds a term quadratic in the energy density, which is familiar in brane cosmology 
 \cite{BDL}, as well as a term which goes like $1/\tilde \rho^2$ and which is characteristic of non $Z_2$-symmetric brane cosmology \cite{Ida}, i.e. when the two bulk regions surrounding the brane have a different geometry.

It is instructive to study the low energy behaviour  of the Friedmann equation by considering an expansion of Eq. (\ref{Friedmann})
around a static brane configuration at $a=a_0$ such
that $\dot{a}|_{a=a_0}=0$. From Eq. (\ref{junction3}), 
the energy density at $a=a_0$ is given by
\begin{eqnarray}
\Sigma_0=-\Big(\frac{3}{a}+\frac{C'}{C}\Big)
\Big(\tilde B_+^{-1}-\tilde B_0^{-1}\Big)
\Big|_{a=a_0}\,,
\end{eqnarray}
for the $(+)$ brane and 
\begin{eqnarray}
\Sigma_0=-\Big(\frac{3}{a}+\frac{C'}{C}\Big)
\Big(\tilde B_0^{-1}-\tilde B_{-}^{-1}\Big)
\Big|_{a=a_0}\,,
\end{eqnarray}
for the $(-)$-brane.
For the $(+)$-brane, we obtain 
\begin{eqnarray}
\frac{\dot{a}^2}{a^2}
\approx \frac{8\pi G_{\rm eff} }{3}\delta\tilde \rho
+O(\delta\tilde \rho^2)\, \label{low_e}
\end{eqnarray}
where the effective gravitational coupling is given by
\begin{eqnarray}
8\pi G_{\rm eff}=\frac{3(1-\alpha)\rho_+}
                      {2\pi (R_0-R_+)\ell M_6^4 a\big(3+\frac{a f_{,a}}{2f}\big)}
\Big|_{a=a_0}
\,.\label{grav_coup}
\end{eqnarray}
We observe that the gravitational coupling depends strongly on the scale factor and is even negative
for $a_0>\rho_+ \left(\frac{3(1-\alpha^8)}{8(1-\alpha^3)}\right)^{1/5}$ ,
namely in the vicinity of the northern pole.
Therefore, one cannot expect the cosmology governed by the Friedmann equation (\ref{Friedmann}) to coincide with  standard cosmology in the low energy limit, in contrast to the 5D Randall-Sundrum cosmology.
The conclusion here does not depend on the choice of  brane matter.
A similar expression for the effective gravitational coupling
for the $(-)$- brane is obtained 
by replacing $R_0$ and $R_+$ in Eq. (\ref{grav_coup})
with, respectively, $R_-$ and $R_0$.

Now, we consider more specifically   the cosmological evolution 
when the brane matter consists of  a complex scalar field.
We must also take into account the junction condition for the gauge field which is given by
\begin{eqnarray}
 {\cal F}(a)\Big[X\Big]
=-
 e\Phi^2
 (\partial_{\theta}\sigma
  -eA_{\theta})\,,
\label{junction4}
\end{eqnarray}
where we have defined the function  ${\cal F}(a)$   
\beq
{\cal F}(a)\equiv \left(\frac{dA}{dw}\right)^{-1}
F_{w\theta}\big|_{A=a}
=-\frac{2 c_I R_I S(\alpha)}
{(1-\alpha)\rho_+}\left(\frac{\rho_+}{a}\right)^4.
\eeq
Let us now combine the four junction conditions above. By 
subtracting Eq (\ref{junction1}) and Eq.(\ref{junction2}), and then 
 using Eqs. (\ref{p_ext}), (\ref{p_int}) and (\ref{junction4}), one gets the expression 
\begin{eqnarray}
\Big[X\Big]
=-e^2\Phi^2 
  \frac{C^2}{{\cal F}^2}\Big(\frac{C'}{C}-\frac{1}{a}\Big)
\, ,\label{X}
\end{eqnarray}
where the right-hand side depends only on $a$ and $\Phi$ but not their derivatives.
By substituting this expression back in the junction condition (\ref{junction3}), one finds
that the energy density of the brane $\Sigma$ is given explicitly in terms of  $a$ and $\Phi$ only:
\begin{eqnarray}
\Sigma&=&
e^2\Phi^2
 \frac{C^2}{{\cal F}^2}
  \Big(\frac{C'}{C}+\frac{3}{a}\Big)
  \Big(\frac{C'}{C}-\frac{1}{a}\Big)
\, 
\nonumber\\
&=&
\frac{e^2\Phi^2(1-\alpha)^2}
       {4S^2}
 \frac{a^8 f(a)}{\rho_+^6}
 \Big(\frac{f'(a)}{2f(a)}-\frac{1}{a}\Big)
 \Big(\frac{f'(a)}{2f(a)}+\frac{3}{a}\Big).\label{ed}
\nonumber\\
\end{eqnarray}
One can even get a more precise constraint on the energy density in the brane by noting that 
the substitution of (\ref{X}) into the gauge field junction condition (\ref{junction4}) yields
\begin{eqnarray}
  \partial_{\theta}\sigma
  -eA_{\theta}
=e\frac{C^2}{\cal F}\big(\frac{C'}{C}-\frac{1}{a}\big).
\end{eqnarray}
This implies that the part of the energy density that  depends only on the scalar field $\Phi$ is 
constrained to be a specific function of $a$ and $\Phi$, namely 
\begin{eqnarray}
&& V(\Phi)
 +\frac{1}{2}\dot{\Phi}^2
=\frac{1}{2}e^2\Phi^2\frac{C^2}{{\cal F}^2}
 \Big(
  \frac{7}{a}
 +\frac{C'}{C}
 \Big) \big(\frac{C'}{C}-\frac{1}{a}\big)
\,.\label{potentialoftheradialscalarmode}
\nonumber  \\
&&
\end{eqnarray}
By subtracting Eq. (\ref{junction1}) from Eq. (\ref{junction3}), we also obtain
\begin{eqnarray}
\Big(\ddot{a}+\frac{\tilde B'}{\tilde B}\dot{a}^2\Big)
\Big[\frac{1}{X}\Big]
=\dot{\Phi}^2
\,
\label{1/X}
\end{eqnarray}
which governs the brane acceleration.

The  bulk  $\alpha<w<1$ can be divided into two regions, where the branes will be confined: a brane in a given region cannot move to the other one. This 
can be seen by noting that 
the sign of 
\beq
[X]_{\pm}=\pm
\Big(\sqrt{\frac{(1-\alpha)^2\rho_+^2 f(a)}{4 R_{\pm}^2}+\dot{a}^2}\ -\ \sqrt{\frac{(1-\alpha)^2\rho_+^2 f(a)}{4 R_0^2}+\dot{a}^2}\Big)
\eeq
 is determined by the relative values of $R_I$ on the two sides of the brane and thus cannot change. From Eq.~(\ref{X}), one sees that the sign of $[X]$ is 
 negative for
$\alpha < w< w_1$ and positive for $w_1<w<1$, where
\beq
w_1:=
\alpha 
\left(\frac{8(1-\alpha^5)}{5(1-\alpha^8)}\right)^{1/3}
\,,\label{z1}
\eeq
is the value
 at which 
 $\frac{C'}{C}-\frac{1}{a}$, which is plotted in Fig. 1, vanishes.
 Consequently, $w=w_1$ represents a limit that the branes cannot cross.

The existence of this critical value is a consequence of the junction
conditions and our choice of brane matter. Indeed, on the one hand, we
always have $p>p_\theta$ with the complex scalar field that was assumed.
On the other hand, the extrinsic curvature tensor is dictated by the
geometry and $w_1$ is the value where the sign of $\hat
K_i^i/3-\hat K_\theta^\theta$ changes. Since the sign of
$p-p_\theta$ cannot change, we see that the brane cannot cross $w_1$
because the junction conditions cannot be satisfied on the other side.
By noting that $R_+< R_-$, which follows from the relations (\ref{l}) and (\ref{c_I}), one finds only two possibilities for the locations of the branes. The first possibility is to put the $(+)$-brane in the northern region, i.e. $w_+ > w_1$, and the $(-)$-brane in the southern region, i.e. $w_-<w_1$. Another possibility is to put both branes in the nothern region, i.e. $w_+>w_->w_1$. It is however impossible to put both branes in the southern region.

Finally, it is worth noticing that
the brane energy density $\Sigma$ is negative  when  $w_1<w< w_2$, where 
\begin{eqnarray}
 w_2:=
\left(\frac{3(1-\alpha^8)}{8(1-\alpha^3)}\right)^{1/5}\,\label{w2}
\end{eqnarray}
is the value at which the function $\frac{C'}{C}+\frac{3}{a}$ vanishes.
Note that, whereas the coefficient in front of $\tilde\rho^2$
in the Friedmann equation Eq. (\ref{Friedmann}) as well as the effective gravitational coupling diverge when 
one approaches $w_2$, the energy density Eq. (\ref{ed})
tends to zero as well so that the overall term does not diverge: the brane 
can cross smoothly $w_2$, simply with a change of sign for the energy density.

\begin{figure} 
\begin{center}
  \begin{minipage}[t]{.45\textwidth}
   \begin{center}
    \includegraphics[scale=.65]{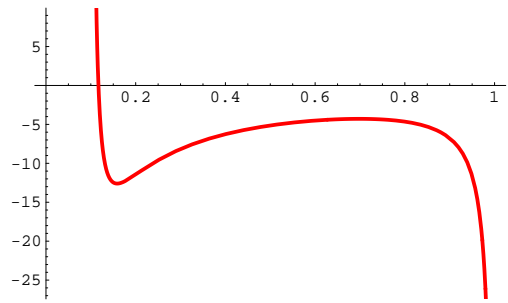}
        \caption{Numerical plot of $C'/C-1/a$ in the Einstein-Maxwell model
 is shown as a function of $a/\rho_+$ for $\alpha=0.1$.}
   \end{center}
   \end{minipage}
   \end{center}
\end{figure}

\subsection{Brane dynamics}

To summarize, we have shown in the previous subsection that, after eliminating
the explicit dependence on the brane Goldstone mode $\sigma$
and vector potential
$A_{\theta}$
via the Maxwell junction condition, the other three junction conditions yield
\begin{eqnarray} 
&&\sqrt{{\tilde B}_2^{-2}+ \dot{a}^2}
-\sqrt{{\tilde B}_1^{-2}+ \dot{a}^2}
 =\Phi^2 G(a)\,,\label{junc1}
\\
&&\frac{1}{\Phi^2}
\Big(
\frac{1}{2}\dot{\Phi}^2
+V(\Phi)
\Big)=K(a)\,,\label{junc2}
\\
&& 
\Big(\ddot{a}
   -\frac{\partial_a f}{2f}
    \dot{a}^2
\Big)
\Big(
\frac{1}{\sqrt{{\tilde B}_2^{-2}+\dot{a}^2}}
-\frac{1}{\sqrt{{\tilde B}_1^{-2}+ \dot{a}^2}}
\Big)
= \dot{\Phi}^2\,,
\nonumber\\ &&
\label{junc3}
\end{eqnarray}
where 
\begin{eqnarray}
&&K(a):=\frac{e^2 (1-\alpha)^2}{8S^2\rho_+^6}
      f(a)a^8
             \Big(\frac{\partial_a f}{2f}
                 +\frac{7}{a}
             \Big)
 \Big(\frac{\partial_a f}{2f}
                 -\frac{1}{a}
             \Big)\,,
\nonumber \\
&&
G(a):=-\frac{e^2 (1-\alpha)^2}{4S^2\rho_+^6}
      f(a)a^8
 \Big(\frac{\partial_a f}{2f}
                 -\frac{1}{a}
             \Big)\,.
\end{eqnarray}
The above equations apply to both branes: for the northern $(+)$-brane,
the subscripts   $"1"$ and $"2"$ correspond to the  $"0"$ and $"+"$ bulk solutions while for the $(-)$-brane,
they correspond to the  $"-"$ and $"0"$ bulk solutions, 
respectively.

The first equation (\ref{junc1}) implies
\begin{eqnarray}
&&\frac{1}{4\Phi^4G^2}\left[\Phi^4G^2-({\tilde B}_1^{-1}+{\tilde B}_2^{-1})^2\right]
\left[\Phi^4G^2-({\tilde B}_1^{-1}-{\tilde B}_2^{-1})^2\right]
\nonumber\\
&&=\dot a^2=: {\cal H}(a,\Phi)^2.
\label{H2}
\end{eqnarray}
By noting that 
\begin{eqnarray}
&&\sqrt{{\tilde B}_1^{-2}+ {\cal H}^2}=-\frac{\Phi^4 G^2+{\tilde B}_1^{-2}-{\tilde B}_2^{-2}}{2\Phi^2 G},
\nonumber\\
&&\sqrt{{\tilde B}_2^{-2}+ {\cal H}^2}=\frac{\Phi^4 G^2+{\tilde B}_2^{-2}-{\tilde B}_1^{-2}}{2\Phi^2 G}
\end{eqnarray}
and that the acceleration of the scale factor can be reexpressed as 
\beq
\ddot a
=\frac{d}{dt}\dot a
=\dot a \frac{d}{da}{\cal H}={\cal H}
\left(
\frac{\partial {\cal H}}{\partial a}
+\frac{\partial {\cal H}}{\partial\Phi}
 \frac{d\Phi}{da}\right),
\eeq
we find,
after some  manipulations, 
 that (\ref{junc3}), divided by $\dot a^2$, can be reduced to
\begin{eqnarray}
&&\Big(\frac{d\Phi}{da}\Big)^2
=\frac{4\Phi^6 G^3}
       {\big(\Phi^4 G^2 - (\tb_1^{-1}+\tb_2^{-1})^2\big)
        \big(\Phi^4 G^2 - (\tb_1^{-1}-\tb_2^{-1})^2\big)}
\nonumber\\
&&\times 
\Big(\frac{2}{\Phi}\frac{d\Phi}{da}
 +\Big(\frac{\partial_a G}{G}-\frac{\partial_a f}{2f}\Big)
\Big)\,
\label{Phieq}.
\end{eqnarray}
This equation can be seen as a first order differential equation for the function $\Phi(a)$.
Given some initial condition $\Phi_{\ast}=\Phi(a_{\ast})$ at some initial point $a_{\ast}$, one can integrate this differential equation and thus obtain $\Phi(a)$. This solution $\Phi(a)$ can then be substituted into (\ref{H2}), from which one can extract
the evolution of the scale factor as a function of the cosmic time $\tau$.
So far, we have not used the second equation (\ref{junc2}). Since the evolution of 
$\Phi$ and  of $a$ can be determined solely from the two other equations, this additional equation plays 
the role of a constraint.
 This means that the potential cannot be arbitrarily chosen but is instead dictated by the geometry. Note that this is a consequence of requiring that the bulk is described by a given static solution. Thus, once the cosmological evolution is known,
 the  potential for the radial scalar field is determined by
\begin{eqnarray}
V=-\frac{1}{2}\cH^2
          \Big(\frac{d\Phi}{da}\Big)^2
         +\Phi^2 K(a)\,,\label{pot}
\end{eqnarray}
which can be expressed in terms of $\Phi$ by inverting the function $\Phi=\Phi(a)$.

\subsection{Behaviour near the "barrier"}

As we saw earlier, $w=w_1$ represents an uncrossable barrier for the branes. It is thus interesting to study
the cosmological behaviour of a brane that approaches this critical value. 
In the limit $a \to a_1:=w_1\rho_+$, the function $G$ vanishes.
Keeping in the differential equation (\ref{Phieq}) the terms that dominate in
this limit, one obtains
\begin{eqnarray}
 \Big(\frac{d\Phi}{da}\Big)^2
&\simeq&
\frac{8\Phi^6 G^3}
{\big(\tb_2^{-2}-\tb_1^{-2}\big)^2}
\Big\{\frac{1}{\Phi}\frac{d\Phi}{da}
    +\frac{1}{2}
    \Big(\frac{\partial_a G}{G}
       - \frac{\partial_a f}{2f} 
    \Big)
\Big\}\,,
\nonumber\\
\label{approxeq}
\end{eqnarray}
where one can substitute $G(a)=G'(a_1)(a-a_1)$.

One can then easily find the approximate solution 
\beq
 \Phi\simeq \Phi_1 + {\tilde \Phi}_1 (a-a_1)^2,\quad
{\tilde\Phi}_1^2=\Phi_1^6\left( \frac{{G'}^3}
              {(\tb_2^{-2}-\tb_1^{-2})^2}\right)_{|a=a_1}\,. 
\label{asympt}
\eeq
The potential Eq. (\ref{pot}) approaches the constant value
\beq
V
\simeq  -\frac{1}{2}\Phi_1^2 G'(a_1) \,,
\eeq
where we have used Eq. (\ref{asympt}).
The above analysis applies for the $(+)$-brane (and possibly for the $(-)$-brane) in the northern region when the brane approaches $a_1$ from above. It is also valid in the southern region when the $(-)$ brane 
 approaches $a_1$ from below (the $(+)$-brane cannot be located in the southern region $a<a_1$ as discussed 
 earlier in subsection B).

Before closing this subsection, we give the scale factor and the scalar field 
in terms of the cosmic time. In the northern region (i.e. $a>a_1$), we find
\begin{eqnarray}
 a-a_1
\simeq
\left(\frac{\sqrt{\tb_2^{-2}-\tb_1^{-2}}}
           {\Phi_1\sqrt{G'}}
\right)_{|a=a_1}
 (\tau_1-\tau )^{1/2}
\end{eqnarray}
where $\tau_1$ the time when the brane reaches $w_1$.
Note that $\tb_2^{-2}> \tb_{1}^{-2}$ in the northern region ($w_1<a/\rho_+< 1$)
because $R_2<R_1$ ($R_+<R_0$ for the $(+)$ brane or $R_0<R_-$ for the $(-)$ brane).

The scalar field evolution is then given by
\begin{eqnarray}
\Phi\simeq
     \Phi_1
    \Big(1
    +\sqrt{G'(a_1)}(\tau_1-\tau)
    \Big)\,.
\end{eqnarray}
This shows that the scalar field velocity approaches a constant and that the ratio $\dot\Phi/\Phi$, in this 
limit, is independent of the initial conditions.
This analysis is similar in the limit
$a\to a_1^-$ on the southern side: one just replaces $a-a_1$ by $a_1-a$ and
$\tb_2^{-2}-\tb_1^{-2}$ by $\tb_1^{-2}-\tb_2^{-2}$.

\subsection{Behavior around the poles}

\subsubsection{$(+)$-brane}

In order to analyse the behaviour of  the differential equation for $\Phi$ , Eq.~(\ref{Phieq}), it
is convenient to introduce the function $C_+(a)$ defined by
\begin{eqnarray}
\Phi^2= C_+^{1/2}f^{1/2} G^{-1}\,.
\end{eqnarray}
If one assumes that  $C_+$ is smooth and non-vanishing in the limit  $a\to\rho_+$, one finds rom
 Eq. (\ref{Phieq})  the following behaviour:
\begin{eqnarray}
\frac{1}{q_+}-C_+
\simeq A_+ \Big(\rho_+ - a\Big)^{\Gamma_+}
\,,\label{approx_sol_p}
\end{eqnarray}
where $A_+$ is an integration constant and where we have introduced the notation
\begin{eqnarray}
&&
\frac{1}{q_+}:= \frac{(1-\alpha)^2\rho_+^2}{4}
    \Big(\frac{1}{R_+}-\frac{1}{R_0}\Big)^2\,,
\nonumber\\
&&  \frac{1}{r_+} := \frac{(1-\alpha)^2\rho_+^2}{4}
    \Big(\frac{1}{R_+}+\frac{1}{R_0}\Big)^2\,,
\label{r_q}
\end{eqnarray}
and
\begin{eqnarray}
\Gamma_+&:=&\frac{1}{32}\frac{-f'(\rho_+)}{G(\rho_+)}
  \big(\frac{1}{r_+}-\frac{1}{q_+}\big)
 =\frac{3}{20}\frac{1}{e^2R_+R_0}
  \frac{\alpha^3(1-\alpha^5)}{1-\alpha^8}
\nonumber\\
&=&\frac{3c_0}{e^2\ell^2}
 \frac{(1-\alpha)\alpha^3(1-\alpha^5)}
      {(5-8\alpha^3+3\alpha^8)}\,,
\end{eqnarray}
using (\ref{l}) in the last equality.

Note that the condition $C_+= 1/q_+$ is equivalent to
the codimension 2 limit
\begin{eqnarray}
  2\pi \ell \sqrt{f(a)} \Sigma_{+}
=\delta_{N}
\,,
\end{eqnarray}
where $\delta_N$ is the deficit angle
at the northern pole $\rho=\rho_+$,
which can be related to the coefficients $c_0$ and $c_+$ \cite{PPZ, KM}
\begin{eqnarray}
 \frac{c_0}{c_+}
=1-\frac{\delta_N}{2\pi}\,
.
\end{eqnarray}
 In order to express the scale factor in terms of the cosmic time $\tau$,
the simplest way is to use
Eq. (\ref{H2}), which yields
\begin{eqnarray}
\dot{a}^2
&=&
\frac{\big(C_+ -\frac{1}{q_+}\big)\big(C_+ - \frac{1}{r_+}\big)} 
     {4C_+} f
\nonumber\\
&\simeq&
\frac{-f'(\rho_+)}{4}q_+\big(\frac{1}{r_+}-\frac{1}{q_+}\big)
A_+ (\rho_+-a)^{\Gamma_+ + 1}
.
\end{eqnarray}
Then, expanding around $C\simeq 1/q_+$ and substituting
the behaviour  Eq. (\ref{approx_sol_p})
into the above expression, we obtain
\begin{eqnarray}
\rho_+-a \simeq B_+(\tau_+-\tau)^{\beta_+} \label{e}
\end{eqnarray}
where
\begin{eqnarray}
&&\beta_+ =\frac{2}{1-\Gamma_+}\,,
\nonumber\\
&&B_+ =
\left(\frac{q_+\big(\frac{1}{r_+}-\frac{1}{q_+}\big) (-f'(\rho_+))A_+}
            {4\beta_+^2}
\right)^{1/(1-\Gamma_+)}\,. \label{u}
\end{eqnarray}
Thus for $\Gamma_+<1$, we obtain $\beta_+ >2$, namely 
the velocity and acceleration are vanishing at the position of the conical
singularity.
 For the $(+)$-brane, we have the condition $c_+>c_0$ ($R_+<R_0$) and thus
\begin{eqnarray}
e^2\ell^2 \Gamma_{+}<
\frac{60\alpha^3(1-\alpha)^2(1-\alpha^3)(1-\alpha^5)}
     {(5-8\alpha^3+3\alpha^8)^2}\leq \frac{1}{4}\,.
\end{eqnarray}
One can thus get  $\Gamma_+<1$ for sufficiently large
$e^2\ell^2$.
Noting that 
\begin{eqnarray}
{\cal H}\simeq \frac{2}{1-\Gamma_+} B_+^{(1-\Gamma_+)/2}
    \Big(\rho_+ - a\Big)^{(1+\Gamma_+)/2}\,, 
\end{eqnarray}
we find that, if $\Gamma_+<1$,
the brane reaches the northern pole in a finite time. 
Indeed, 
\begin{eqnarray}
(\Delta\tau)_+= \int_{a_{\ast}}^{\rho_+}
    \frac{da}{{\cal H}(a,\Phi(a))}\,,
\end{eqnarray}
is finite if $(1+\Gamma_+)/2<1$ ($a_{\ast}$
is the initial position of the brane).
Thus, it is impossible to obtain an ever expanding brane Universe.


\subsubsection{$(-)$-brane}

The analysis of the opposite limit  $a\to \rho_-=\alpha\rho_+$ is quite similar to the previous case.
In the region $\rho_- <a<w_1\rho_+$, we always have
$G(a)<0$, and it is now convenient to introduce $C_-(a)$ defined by 
\begin{eqnarray}
\Phi^2= C_-(a)^{1/2}f^{1/2} (-G)^{-1}\,.
\end{eqnarray}
Assuming $C_-$ is non-vanishing and regular in the limit  $a= \rho_-$, one can easily find the 
behaviour
\begin{eqnarray}
\frac{1}{q_-}-C_-\simeq A_-
  \big(a-\rho_-\big)^{\Gamma_-}
\end{eqnarray}
where $A_-$ is an integration constant and  
\begin{eqnarray}
\Gamma_-&:=&\frac{1}{32}\frac{f'(\rho_+\alpha)}{-G(\rho_+\alpha)}
  \big(\frac{1}{r_-}-\frac{1}{q_-}\big)
 =\frac{3}{20}\frac{1}{e^2 R_-R_0}
  \frac{1-\alpha^5}{\alpha^5 (1-\alpha^8)}
\nonumber\\
&=&\frac{3c_0}{e^2\ell^2}
\frac{(1-\alpha)(1-\alpha^3)}
     {\alpha(5\alpha^8-8\alpha^5+3)},
\end{eqnarray}
with $q_-$ and $r_-$ defined as in (\ref{r_q}) by replacing $R_+$ by $R_-$.

Note that the condition $C_-= 1/q_-$ is equivalent to
the conical limit
\begin{eqnarray}
  2\pi \ell \sqrt{f(a)} \Sigma_{-}
=\delta_{S}
\,,
\end{eqnarray}
where $\delta_S$ is the deficit angle at the southern pole
and we have used the relation \cite{PPZ,KM}
\begin{eqnarray}
 \frac{c_0}{c_-}
=1-\frac{\delta_S}{2\pi}\,.
\end{eqnarray}
The expression of the scale factor in terms of the cosmic time is  given by
\begin{eqnarray}
a-\rho_-  \simeq B_-(\tau_- - \tau)^{\beta_-}
\end{eqnarray}
with
\begin{eqnarray}
&&\beta_- =\frac{2}{1-\Gamma_-}\,,
\nonumber\\
&&B_- =
\left(\frac{q_-\big(\frac{1}{r_-}-\frac{1}{q_-}\big) (f'(\rho_-))A_-}
            {4\beta_-^2}
\right)^{1/(1-\Gamma_-)}\,,
\end{eqnarray}
and we find, as near the northern conical singularity, that the brane approache
$\rho_-$ in a finite time if $\Gamma_-<1$.

\subsection{Numerical example}

In order to illustrate the cosmological behaviour of the branes and to confirm our analytical estimates in the various limits discussed above, we have integrated numerically (\ref{Phieq}) for the $(+)$-brane 
 by starting from some initial condition $\Phi_*=\Phi(a_*)$.
Our choice of parameters is the following:
\beq
\alpha=0.1, \quad 
e\rho_+=1, \quad R_+/\rho_+=4.0, \quad 
R_0/\rho_+=5.0
\eeq
This gives $w_1=0.11696$.
The initial position and initial amplitude are set to be
\begin{eqnarray}
a_{\ast}= 0.2\rho_+ \,, \quad 
\Phi=0.01 \frac{M_6^2}{e \sqrt{\rho_+}}\,. 
\end{eqnarray}
From this initial position, we have integrated for decreasing values of $a$ until reaching $w_1$ as well as for 
increasing values of $a$ up to the northern conical singularity.


\begin{figure}
\begin{center}
  \begin{minipage}[t]{.45\textwidth}
   \begin{center}
    \includegraphics[scale=.65]{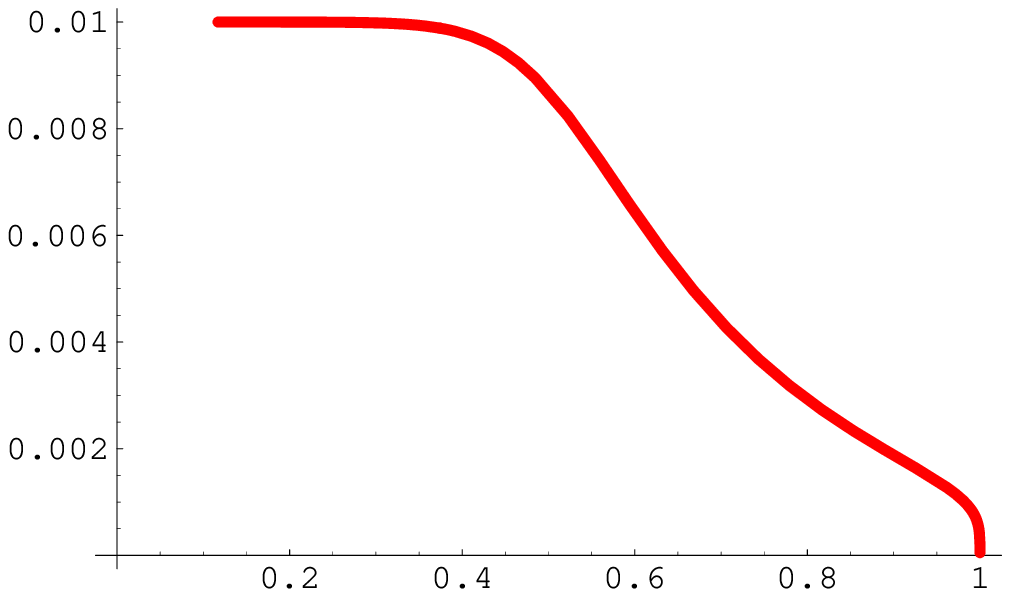}
        \caption{Numerical plot of $\Phi(a)$ in the Einstein-Maxwell model, with the parameters 
        $\alpha=0.1$, $e\rho_+=1$, $R_+/\rho_+=4.0$, $R_0/\rho_+=5.0$. $\Phi(a)$
 is shown as a function of $a$ (in units of $\rho_+$) for  an initial amplitude 
 $\Phi_{\ast}=0.01 \frac{M_6^2}{e \sqrt{\rho_+}}$ at the initial position
$a_{\ast}= 0.2\rho_+$. 
}
   \end{center}
   \end{minipage}
   \hspace{0.5cm}
   \begin{minipage}[t]{.45\textwidth}
   \begin{center}
    \includegraphics[scale=.65]{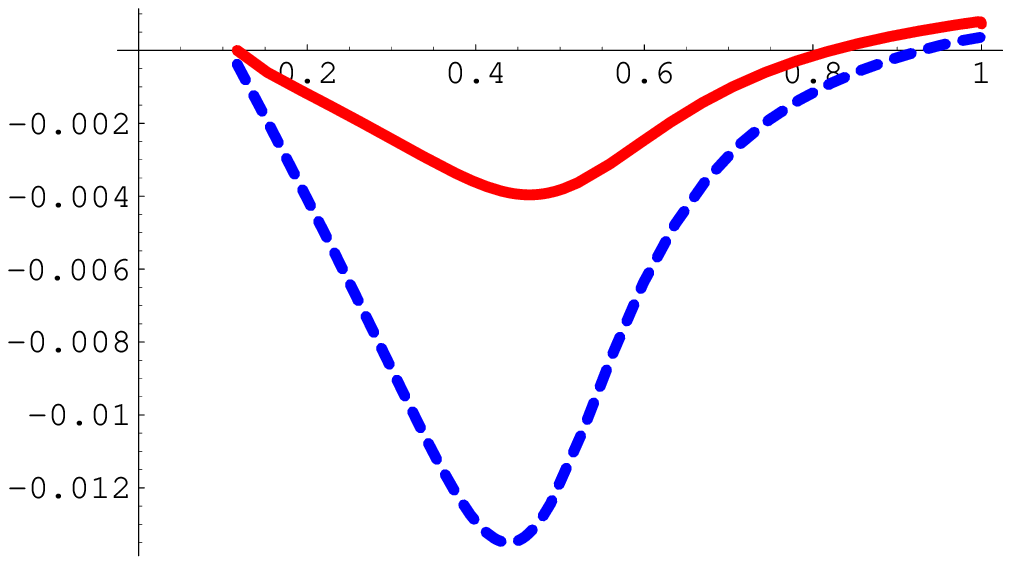}
\caption{Numerical plots of $\sqrt{f} \, V(\Phi)$ (blue, dashed curve)
and $\sqrt{f}\, \Sigma$ (red, solid curve) are shown 
as functions of $a$ (in the unit of $\rho_+$) for the solution of Fig. 2. 
Note that the total energy density is vanishing at $a=\rho_+ w_1$ while
the potential $V(\Phi)$ is negative at the same point.}
   \end{center}
   \end{minipage}
   \end{center}
\end{figure}
\begin{figure}
\begin{center}
  \begin{minipage}[t]{.45\textwidth}
   \begin{center}
    \includegraphics[scale=.65]{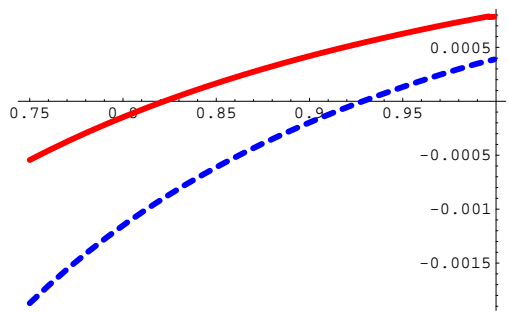}
        \caption{Enlargement of Fig. 3 near
the northern pole (conical singularity) $a=\rho_+$. 
As we see, the relation $\Sigma= 2V$ is satisfied at the conical singularity
limit.}
   \end{center}
   \end{minipage}
   \end{center}
\end{figure}


We have plotted the profile of $\Phi(a)$ in Fig. 2. We observe that the scalar field vanishes at the conical
singularity and approaches a constant value, with vanishing speed, at $w=w_1$, which is consistent with our analytical solutions. 

In Fig. 3 we have plotted the effective potential and total energy densities, respectively $\sqrt{f} V$ and
$\sqrt{f}\Sigma$. 
Near the conical singularity, as shown in Fig. 4,
we see that the special relation
$\Sigma= 2V$ is satisfied, which corresponds to the conical
singularity condition.

Our numerical example corresponds to the case where the brane reaches the conical singularity
in a finite time as discussed earlier.


\section{Einstein-Maxwell-dilaton model}

\subsection{Bulk spacetime in 6D Einstein-Maxwell-dilaton model}

In this section, we consider a slightly different model based on solutions of the 6D Einstein-Maxwell-dilaton theory, which now include a dilaton denoted $\varphi$. Our bulk action is based 
on the bosonic part
of the Nishino-Sezgin supergravity \cite{6d_sugra}:
\begin{eqnarray}
S_{B}&=&\int d^6 x\sqrt{-g}
\Big(
 \frac{1}{2}R
-\frac{1}{2}(\partial_A \varphi)^2
-\frac{1}{4}e^{-\varphi}F_{AB}F^{AB}
\nonumber\\
&-&4g_0^2 e^{\varphi}
\Big) \,,
\end{eqnarray} 
where the other fields can be consistently set to zero.
It is possible to find exact solutions, given by \cite{6d_sugra}
\begin{eqnarray}
ds^2&=&(2\rho_+ w)\eta_{\mu\nu}dx^{\mu}dx^{\nu}
\nonumber\\
&+&\frac{\rho_+}{2g_0^2}
\Big(\frac{4dw^2}{(1-\alpha)^2f(w)}
    +f(w) c_0^2 d\theta^2
\Big)\,,\label{metric_emd1}
\\
 F_{w\theta}
&=&-\frac{\sqrt{2}c_0 \alpha}{(1-\alpha) g_0 w^3}\,,\quad
\varphi= -\ln(2\rho_+ w)\,,
\label{F_emd1}
\end{eqnarray}
where
\begin{eqnarray}
f(w)=\frac{2(1-w^2)(w^2-\alpha^2)}
              {w^3(1-\alpha)^2}\,
.
\end{eqnarray}
The geometry possesses two  conical singularities at $w=\alpha$ and $w=1$, respectively. As in the non-dilatonic case, we regularize the 
corresponding codimension 2 branes by introducing  two codimension-1 branes at the respective positions $w_->\alpha$ and $w_+<1$, delimiting two regular  caps. The northern cap  ($w_+<w<1$) and the southern cap 
($\alpha<w<w_-$), can be described by solutions similar to that of the main bulk, i.e. 
of the form (\ref{metric_emd1})-(\ref{F_emd1}), but with parameters $g_{\pm}$ and $c_\pm$ that differ from 
$g_0$ and $c_0$.
The regularity of the caps imposes \cite{PPZ}
\begin{eqnarray}
c_{+}=\frac{1}{1+\alpha}\,,\quad
c_{-}=\frac{\alpha^2}{1+\alpha}\,,
\end{eqnarray}
while 
the continuity of the $(\theta,\theta)$ component requires
\begin{eqnarray}
\ell^2= \frac{c_+^2}{2g_+^2}=\frac{c_-^2}{2g_-^2}
=\frac{c_0^2}{2g_0^2}\,.
\end{eqnarray}
The solutions for the bulk gauge field around the north and
south poles are given by
\begin{eqnarray}
A^{(N)}_{\theta}=\frac{\alpha \ell}{1-\alpha}
        \big(\frac{1}{w^2}-1\big)\,,\quad
A^{(S)}_{\theta}=\frac{\alpha \ell}{1-\alpha}
        \big(\frac{1}{w^2}-\frac{1}{\alpha^2}\big)\,.
\end{eqnarray}
The difference should be a pure gauge term,
which implies the Dirac quantization condition
\begin{eqnarray}
N=\frac{e\ell (1+\alpha)}{\alpha}\,.
\end{eqnarray}

\subsection{Junction conditions and brane dynamics}

We describe the brane matter by an  action of the form
\begin{eqnarray}
S_{b,\pm}&=&-\int d^5 x\sqrt{-q}
    \Big[V(\Phi_{\pm},\varphi)
    +\frac{1}{2}\xi(\varphi)
    \Big((\partial_a \Phi_{\pm})^2
\nonumber\\
&   +&\Phi_{\pm}^2\Big(\partial_a \sigma_{\pm}-eA_a\Big)^2
\Big)
    \Big]\,.
\end{eqnarray}
It is similar to the non-dilatonic case with two differences: 
the potential $V(\Phi_{\pm}, \varphi)$ now depends on the dilaton, and 
we have introduced a coupling $\xi(\varphi)$, assumed to be strictly positive, between  the kinetic term 
of the scalar field and the bulk dilaton.
The energy density $\Sigma$ and the pressures along the external and internal
directions, respectively $p$ and $p_\theta$, are defined like in (\ref{Sigma})-(\ref{p_int}), with the only difference that the potential now depends on the dilaton $\varphi$, in addition to the scalar field $\Phi$.
The comparison of  the metric (\ref{metric_emd1}) with Eq. (\ref{metric_ge}) yields the identification
\begin{eqnarray}
&&A^2=2\rho_+ w\,,\quad
B^2=\frac{2\rho_+ }{g_I^2 (1-\alpha)^2 f(w)}\,,\quad
\nonumber\\
&&C^2=\frac{\rho_+}{2g_I^2}c_I^2 f(w)
  =\rho_+ \ell^2 f(w)
\,,
\end{eqnarray}
and the scale factor of the brane is, as before,
$a(\tau)\equiv A(w_b(\tau))$.
As in the previous section, we  also define the rescaled metric component
\begin{eqnarray}
\tilde B(A)=B\frac{dw}{dA}
=\frac{(2\rho_+)^{1/2}\sqrt{f(a)}}{g_I a (1-\alpha)}
\,.
\end{eqnarray}

The junction conditions for the metric are exactly the same as in the previous section, and are thus given by 
(\ref{Sigma}-\ref{p_int}). They imply in particular 
the generalized Friedmann equation
\begin{eqnarray}
&&\frac{\dot{a}^2}{a^2}
=\frac{{\tilde \rho}^2}
{16\pi^2\ell^2\rho_+ f \big(3+\frac{f'}{2f}a\big)^2} 
\nonumber\\
&&+\frac{\pi^2\ell^2 f}{a^4}
    \frac{\rho_+\big(3+\frac{f'}{2f}a\big)^2 
        (\tilde B{}_{\pm}^{-2}-\tilde B{}_0^{-2})^2}
          {{\tilde \rho}^2}
-\frac{\tilde B_{\pm}^{-2}+\tilde B_0^{-2}}{2a^2},
\nonumber\\
&&\label{Friedmann_s}
\end{eqnarray}
where $\tilde \rho:= 2\pi\ell\sqrt{\rho_+} \sqrt{f}\Sigma$.
The Friedmann equation is very similar to the one obtained in the
 Einstein-Maxwell case and one would find, by repeating the same procedure as in the previous section, that the low-energy behaviour differs from standard cosmology.

The Maxwell junction condition, however, is modified because the dilaton is explicitly coupled to the electromagnetic field.
It now reads
\begin{eqnarray}
 e ^{-\varphi}{\cal F}(a)\Big[X\Big]
=-
e \xi(\varphi) \Phi^2
 (\partial_{\theta}\sigma
  -eA_{\theta})\,,
\label{junction4a}
\end{eqnarray}
where we have defined
\begin{eqnarray}
{\cal F} (a):=
 \left( \frac{dA}{dw}\right)^{-1}
              F_{w\theta}
 =-\frac{16\alpha\ell}{1-\alpha} 
   \frac{\rho_+^2}{a^5}
   \,.
\end{eqnarray}
Moreover, we have an additional junction condition, that of the dilaton,
which is given by 
\begin{eqnarray}
 &&\tilde \varphi(a) \Big[X\Big]
\nonumber \\
&=&
\frac{\partial V}
 {\partial\varphi}
+\frac{1}{2}\xi'(\varphi)
\Big( 
  - \dot{ \Phi}^2
   +\Phi^2 C^{-2}\Big(\partial_{\theta} \sigma
                     -eA_{\theta}\Big)^2
\Big)
\,,
\label{junction5a}
\nonumber\\
&&
 \tilde \varphi(a):=\left( \frac{dA}{dw}\right)^{-1}
                       \varphi_{,w}
=-\frac{2}{a}\,.
\end{eqnarray}
As in the Einstein-Maxwell model, the bulk is divided in two by a limit that cannot be crossed by the branes. 
In the present case, the position of this limit is given by
\begin{eqnarray}
w_1:=\sqrt{\frac{2}{1+\alpha^2}}\alpha\,.
\label{app_emd}
\end{eqnarray}
Similarly, the total energy density in the brane is negative when 
\begin{eqnarray}
 w_1
<\frac{a^2}{2\rho_+}
<w_2:=\sqrt{\frac{1+\alpha^2}{2}}\,,
\end{eqnarray}
and positive otherwise.
As in the case of the Einstein-Maxwell model, the $(+)$-brane 
cannnot be in the region $(2\rho_-)^{1/2}<a<(2\rho_+w_1)^{1/2}$, where $\rho_-:=\rho_+\alpha$.
 The brane can also cross $w_2$ simply with a change of the sign of
the energy density.

Following the same method  as in the Einstein-Maxwell case,
one can rewrite the junction conditions as 
\begin{eqnarray} 
&&\sqrt{\tb_2^{-2}+ \dot{a}^2}
-\sqrt{\tb_1^{-2}+ \dot{a}^2}
 =\xi(\varphi) \Phi^2 G(a)\,,\label{junc1a}
\\
&&\frac{1}{\Phi^2}
\Big(
\frac{1}{2}\xi(\varphi)\dot{\Phi}^2
+V(\Phi, \varphi)
\Big)=\xi(\varphi)K(a)\,,\label{junc2a}
\\
&& 
\Big(\ddot{a}
   +\partial_a\ln(\frac{a}{\sqrt{f}})
    \dot{a}^2
\Big)
\Big(
\frac{1}{\sqrt{\tb_2^{-2}+\dot{a}^2}}
-\frac{1}{\sqrt{\tb_1^{-2}+ \dot{a}^2}}
\Big)
\nonumber\\
&&= \xi(\varphi) \dot{\Phi}^2\,,
\label{junc3a}
\end{eqnarray}
where
\begin{eqnarray}
&&K(a):=\frac{e^2 (1-\alpha)^2}{64\alpha^2(2\rho_+)^3}
      f(a)a^6 
             \Big(\frac{\partial_a f}{2f}
                 +\frac{7}{a}
             \Big)
 \Big(\frac{\partial_a f}{2f}
                 -\frac{1}{a}
             \Big)\,,
\nonumber\\
&&G(a):=-\frac{e^2 (1-\alpha)^2}{32\alpha^2(2\rho_+)^3}
      f(a)a^6
 \Big(\frac{\partial_a f}{2f}
                 -\frac{1}{a}
             \Big)\,.
\end{eqnarray}

This implies in particular 
\begin{eqnarray}
&& V(\Phi,\varphi)
 +\frac{1}{2}\xi(\varphi)\dot{\Phi}{}^2
\nonumber\\
&=&\frac{1}{2}\xi(\varphi)
 \Big(
  \frac{7}{a}
 +\frac{C'}{C}
 \Big)
\frac{e^2\Phi^2C^2\big(\frac{C'}{C}-\frac{1}{a}\big)}
     {e^{-2\varphi} {\cal F}^2}
\,.\label{potentialoftheradialscalarmodea}
\end{eqnarray}
One can also note that the topological conditions are the same as in the static case\cite{PST} 
(see Appendix A-2 for details).

To obtain the cosmological evolution, 
one can proceed exactly as in the previous section. In particular, the square of the scale factor velocity is 
given by
\begin{widetext}
\beq
\dot a^2
= \frac{1}{4\Phi^4G^2\xi(\varphi)^2}
\left(\Phi^4G^2\xi(\varphi)^2-(\tb_1^{-1}+\tb_2^{-1})^2\right)
\left(\Phi^4G^2\xi(\varphi)^2-(\tb_1^{-1}-\tb_2^{-1})^2\right)
:= \cH(a,\Phi(a))^2\,,
\label{cHa}
\eeq
and the differential equation for $\Phi$ is now 
\begin{eqnarray}
\Big(\frac{d\Phi}{da}\Big)^2
=\frac{4\Phi^6 G^3 \xi(\varphi)^3}
       {\big(\Phi^4 G^2\xi(\varphi)^2 - (\tb_1^{-1}+\tb_2^{-1})^2\big)
        \big(\Phi^4 G^2\xi(\varphi)^2 - (\tb_1^{-1}-\tb_2^{-1})^2\big)}
\times\Big(\frac{2}{\Phi}\frac{d\Phi}{da}
 +\Big(\frac{\partial_a G}{G}
      +\partial_a  \ln\big(\frac{a\xi(\varphi)}{\sqrt{f}}\big)\Big)
\Big)\,.
\label{eom_pa}
\end{eqnarray}
\end{widetext}
The junction condition (\ref{junc2a}) implies 
\begin{eqnarray}
V=\xi(\varphi)\Big(-\frac{1}{2}\cH^2
          \Big(\frac{d\Phi}{da}\Big)^2
         +\Phi^2 K(a)
\Big)
\,,\label{potential}
\end{eqnarray}
which provides an expression for the potential $V$ as a function of $a$. The potential 
depends on both the scalar field $\Phi$ and the dilaton $\varphi$, and it is possible to disentangle the dependence on $\Phi$ and $\varphi$ by using
 the dilaton junction condition Eq. (\ref{junction5a}) with
Eq. (\ref{junc1a}), which yields the following expression for the 
partial derivative of the potential with respect to the dilaton:
\begin{eqnarray}
\frac{\partial V}{\partial \varphi}
&=&-
\Phi(a)^2G(a)
\Big(\frac{2}{a}\xi(\varphi(a))
   -
\frac{1}{2}\xi'(\varphi(a))
   \big(\frac{f'}{2f}-\frac{1}{a}\big)
\Big)
\nonumber\\
&+&\frac{1}{2}\xi'(\varphi)
  \Big(\frac{d\Phi}{da}\Big)^2
  {\cal H}(a,\Phi(a))^2\, .
\end{eqnarray}

\subsection{Behaviour near the "barrier"}

We now analyze the behavior near the critical point
$a\to a_1:=(2\rho_+w_1)^{1/2}$.
The differential equation of the scalar field Eq. (\ref{eom_pa})
becomes in the limit $\Phi^4 G^2\xi^2\to 0$
\begin{eqnarray}
&&\xi(\varphi)
\Big(\frac{d\Phi}{da}\Big)^2
\nonumber\\
&&\simeq
\frac{8\Phi^6 G^3\xi^3}
{\big(\tb_2^{-2}-\tb_1^{-2}\big)^2}
\Big\{\frac{1}{\Phi}\frac{d\Phi}{da}
    +\frac{1}{2}
    \Big(\frac{\partial_a G}{G}
      +\partial_a \ln\Big(\frac{a\xi}{\sqrt{f}}\Big)
    \Big)
\Big\}
\,.
\nonumber\\
\label{approxa}
\end{eqnarray}
We expand $G(a)=G'(a_1)(a-a_1)$
and
$K(a)=K'(a_1)(a-a_1)$, with
$G'(a_1)>0$ and $K'(a_1)<0$.
Then, assuming that $\xi$ tends smoothly towards a finite value $\xi_1:=\xi(\varphi(a_1))$, we 
find the following approximate behaviour
\begin{eqnarray}
\Phi\simeq\Phi_0+\Phi_1 (a-a_1)^{2}\,,\quad
\Phi_1^2=\frac{\xi(\varphi)^2 \Phi_0^6 {G'}^3}
              {(\tb_2^{-2}-\tb_1^{-2})^2} \Big|_{a=a_1} \,. 
\label{asympta}
\end{eqnarray}
The potential is given by
\begin{eqnarray}
V \simeq  -\frac{1}{2}\Phi_0^2\xi(\varphi)
 G' \Big|_{a=a_1} <0\,,
\end{eqnarray}
where we have used Eq. (\ref{asympta}).
Thus, the potential is finite and negative in the limit $a\to a_1$.

We now derive the expression for the scale factor 
in terms of the brane proper time.
In the case $a\to a_1^+$,
we obtain
\begin{eqnarray}
a-(2\rho_+w_1)^{1/2}
=\frac{\sqrt{\tb_2^{-2}-\tb_1^{-2}}}
      {\Phi_0\sqrt{G'\xi(\varphi)}}\Big|_{a=(2\rho_+w_1)^{1/2}}
 (\tau_1-\tau )^{1/2}
\nonumber\\
\end{eqnarray}
where $\tau_1$ represents the time when the brane reaches $a_1$.
Note that $\tb_2^{-2}> \tb_{1}^{-2}$ for $a_1
<a< (2\rho_+ )^{1/2}$ because $g_2>g_1$ ($g_+>g_0$).
The scalar field configuration is given by
\begin{eqnarray}
\Phi=\Phi_0 
    \Big(1
    +\sqrt{G'\xi(\varphi)}\Big|_{a=(2\rho_+w_1)^{1/2}}(\tau_1-\tau)
    \Big)\,.
\end{eqnarray}
Thus the scalar field has a constant velocity.
The brane terminates its motion at the point.
The analysis
$a\to a_1^-$
can be done in the similar manner:
one just replaces $a-a_1$
by $a_1-a$ and
$\tb_2^{-2}-\tb_1^{-2}$ by $\tb_1^{-2}-\tb_2^{-2}$.


\subsection{Behaviour around the poles}

In the limit $a\to (2\rho_+)^{1/2}$, it is useful to 
rewrite the scalar field as 
\begin{eqnarray}
\Phi^2= C_+(a)^{1/2}\frac{f^{1/2}}{a\xi(\varphi)} G^{-1}\,.
\end{eqnarray}
If $C_+(a)$ is a smooth and non-vanishing function 
near $a=(2\rho_+)^{1/2}$, an approximate 
solution of the differential equation is given by
\begin{eqnarray}
\frac{1}{q_+}-C_+
\simeq A_+ ((2\rho_+)^{1/2}-a)^{\Gamma_+}
\,,\label{approx_sol_pa}
\end{eqnarray}
where
\begin{eqnarray}
&&\frac{1}{q_+}:= \frac{(1-\alpha)^2\rho_+}{2}
    \big(g_+ - g_0\big)^2\,,
\nonumber\\
&&  \frac{1}{r_+} := \frac{(1-\alpha)^2\rho_+}{2}
    \big(g_+ + g_0\big)^2\,,
\label{q_r_d}
\end{eqnarray}
and
\begin{eqnarray}
\Gamma_+=\frac{1}{32}\frac{-f'((2\rho_+)^{1/2})}
                 {a^2 \xi G\big|_{(2\rho_+)^{1/2}}}
  \Big(\frac{1}{r_+}-\frac{1}{q_+}\Big)
 = \frac{\alpha^2}{e^2\xi\ell^2}\frac{c_0}{1+\alpha}
\,.
\nonumber\\
\end{eqnarray}
Note that the condition $C_+= 1/q_+$ is equivalent to
the conical limit
\begin{eqnarray}
2\pi \sqrt{\rho_+}\ell \sqrt{f(a)}
\Sigma_{+}
\Big|_{a_+\to \sqrt{2\rho_+}}
=\delta_N\,,
\end{eqnarray}
 where $\delta_N$ represents the deficit angle
at the northern pole.
The expression for the scale factor
in terms of $\tau$ is given by
\begin{eqnarray}
(2\rho_+)^{1/2}-a
\simeq B_+ \big(\tau_+ - \tau\big)^{\beta_+}
 \,,
\end{eqnarray}
where
\begin{eqnarray}
&&\beta_+=\frac{2}{1-\Gamma_+}\,,\,
\nonumber \\
&&B_+=\left(
  \frac{q_+\big(\frac{1}{r_+}-\frac{1}{q_+}\big)A_+}
       {4\beta_+^2}
 \frac{-f'((2\rho_+)^{1/2})}{(2\rho_+)}
  \right)^{1/(1-\Gamma_+)}
\end{eqnarray}
For the $(+)$-brane, we have the condition $c_+>c_0$ ($g_+>g_0$) and a discussion similar 
to that of the Einstein-Maxwell model tells us that it is possible, with a sufficiently large 
charge for the brane, to  have $\Gamma_+<1$, in which case the brane reaches the northern pole in a finite
time.
Thus, one does find an ever expanding brane Universe.



The results for the opposite limit  $a\to (2\rho_-)^{1/2}$ are quite similar to the above results.
Using the decomposition,
\begin{eqnarray}
\Phi^2= C_-(a)^{1/2}\frac{f^{1/2}}{a\xi(\varphi)} (-G)^{-1}\,,
\end{eqnarray}
one can identify the behaviour
\begin{eqnarray}
 \frac{1}{q_-}- C_-
\simeq  A_- \Big(a-(2\rho_-)^{1/2}\Big)^{\Gamma_-}
\end{eqnarray}
where
\begin{eqnarray}
\Gamma_-=\frac{1}{32}\frac{f'((2\rho_-)^{1/2})}
                 {-a^2 \xi G\big|_{(2\rho_-)^{1/2}}}
  \Big(\frac{1}{r_-}-\frac{1}{q_-}\Big)
=\frac{1}{e^2\xi\ell^2}\frac{c_0}{1+\alpha}\,
\nonumber\\
\end{eqnarray}
and $q_-$ and $r_-$ are defined as in (\ref{q_r_d}) by replacing $g_+$ with $g_-$.

Note that the condition $C_-=1/q_-$ corresponds to
\begin{eqnarray}
2\pi \sqrt{\rho_+}\ell \sqrt{f(a)}\Sigma_{-}
\Big|_{a_-\to \sqrt{2\rho_-}}
=\delta_S\,,
\end{eqnarray}
 where $\delta_S$ denotes the deficit angle
at the southern pole.
In terms of 
the cosmic proper time $\tau$, we find
\begin{eqnarray}
a-(2\rho_-)^{1/2}
=B_- \big(\tau_{-}-\tau\big)^{\beta_-}
 \,,
\end{eqnarray}
where
\begin{eqnarray}
&&\beta_-=\frac{2}{1-\Gamma_-}\,,
\nonumber\\
&&B_-=\left(
  \frac{q_+\big(\frac{1}{r_-}-\frac{1}{q_-}\big)A_-}
       {4\beta_-^2}
 \frac{f'((2\rho_-)^{1/2})}{(2\rho_-)}
  \right)^{1/(1-\Gamma_-)}\,.
\end{eqnarray}
For the $(-)$-brane, $c_- > c_0$ ($g_->g_0$) and one can have $\Gamma_-<1$, in which case the brane reaches the south pole in a finite time.


\section{Discussions}

In this work, we have investigated the
dynamics of the regularized branes in a 6D bulk.
We considered  models based on either the Einstein-Maxwell theory
or Einstein-Maxwell-dilaton theory (more precisely a bosonic part
of Nishino-Sezgin supergravity).
In both  models, the original systems are composed of a warped bulk
 bounded by codimension 2 branes located at the poles.
In order to introduce non-trivial matter on the branes
we have regularized the codimension 2 branes, located at the poles,
by replacing them with ring-like codimension 1 branes.

The brane matter is composed of a complex scalar field coupled to the bulk $U(1)$ gauge field (and to the bulk dilaton in the dilatonic model).
Such matter was used for a static (fixed) brane, where the radial mode of 
the complex scalar field was assumed to be stabilized at a local minimum of its
potential and thus frozen at a fixed vacuum expectation value  \cite{PST}. 
For moving branes, one must allow for a time dependence of the radial mode
and this radial mode controls the dynamics of the brane in the bulk.

We have studied the dynamics of such branes  in the 
two types of models mentioned above.
In both  models, branes  exhibit very similar behaviors and their 
 essential features can be summarized as follows.
First, there is a critical radius, translated into a critical scale factor $a_1$  given by
$w_1\rho_+$ in the Einstein-Maxwell model (see Eq. (\ref{z1})) and
$\sqrt{2\rho_+ w_1}$ in the Einstein-Maxwell-dilaton model
(see Eq. (\ref{app_emd})), which cannot be crossed by a brane:
an expanding brane with $a<a_1$ at some time cannot expand  beyond $a_1$; conversely, a contracting brane
with $a>a_1$  cannot shrink below $a_1$.
Investigating the motion of a brane close to this critical  point, we have found
 that a brane reaches the point $a_1$, within a finite time, with a divergent velocity
 ($\dot a\propto |\tau-\tau_1|^{-1/2}$, where $\tau$ is the 
proper time on the brane).
Therefore, the Hubble parameter of the brane diverges at $a=a_1$.
Second, near the poles,
we have found behaviours where the brane 
reaches the conical singularity within a finite time
 with vanishing velocity and acceleration. This occurs if the brane charge is sufficienly large.

For the Einstein-Maxwell model, we have solved the brane motion numerically
and  confirmed our analytic results in both limits.
In both types of models, the size of the internal dimension goes to zero as the brane approaches 
a conical singularity or to a constant as the brane approaches the critical value $a_1$. 
These conclusions were obtained by assuming a complex scalar field as brane matter and 
it would be worthwhile investigating the cosmological behaviour  for other kinds of matter.

More generally, i.e. independently of the specific brane matter content,  we have seen that  
the Friedmann equations Eq. (\ref{Friedmann})
and (\ref{Friedmann_s}), in the Einstein-Maxwell and Einstein-Maxwell-dilaton
models, respectively, do not behave like standard cosmology in the low energy limit,
because the effective gravitational coupling, Eq. (\ref{grav_coup})
 is  strongly time-dependent (and even negative in the vicinity of the pole).
Thus, our results show that, in contrast
with the 5D Randall-Sundrum cosmology in the low-energy limit, 
the dynamics of the brane in the present setup
 is not compatible with the usual cosmological evolution 
and modifications are necessary to obtain a sensible cosmology.
One possible modification would be to consider time-dependent bulk solutions 
 \cite{6dtimedep} instead of the static geometries considered here.
\footnote{Low energy behaviour of brane gravity in the Einstein-Maxwell model
was discussed recently in Ref. \cite{Fuji}}.
We will report these investigations in future publications.

\section{Addendum}
While this paper was being completed, we became aware of a related work \cite{PPZ2}, which  studies 
the cosmological behaviour in the same type of models as those discussed here.  As far as we can see, our results agree with those of \cite{PPZ2}. The two works are also complementary in many respects.

\section*{Acknowlegements}
We would like to thank A.
Papazoglou for instructive discussions in the latest stage of this work. 
DL and MM were partially supported by the CNRS-JSPS exchange programme, and 
the work of MM was also supported in part by the project "Dark Universe"
at the ASC.
MM is  grateful for the warm hospitality at IAP and APC. 
DL would like to thank the Yukawa Institute for Theoretical Physics,
Kyoto University,
for their warm hospitality.

\appendix

\section{Brane equations and quantum numbers}

\subsection{Einstein-Maxwell model}

The dynamics of the radial scalar field on the brane is given by
\begin{eqnarray}
 &&\ddot{\Phi}
+\dot{A}\Big(\frac{3}{A}+\frac{C'(A)}{C}\Big)
 \dot{\Phi}
+V_{\Phi}
+\frac{1}{C^2}
 \Big(\partial_{\theta} \sigma
-eA_{\theta} \Big)^2\Phi
\nonumber \\
&=&0\,.
\end{eqnarray}
The equation of motion of $\Phi$ field is consistent with the energy momentum conservation law on the brane.
\begin{eqnarray}
 \dot{\Sigma}
+\frac{3\dot{A}}{A}
 \Big(\Sigma+p\Big)
+\frac{C' \dot{A}}{C}
\Big(\Sigma+p_{\theta}\Big)
=\Big[{\cal T}_{AB}u^A n^B\Big]\,.
\end{eqnarray}
with
\begin{eqnarray}
&& \Big[{\cal T}_{AB}u^A n^B\Big]
=\dot{A}\Big(F_{A\theta}\Big)^2
 \Big[X\Big]C^{-2}
\nonumber\\
&=&\dot{A}C^{-2}
 e\Phi^2 
\big(\partial_{\theta}\sigma
-eA_{\theta}\big)
 F_{A\theta}\,,
\end{eqnarray}
where, in the final step, we have used the Maxwell junction condition.

The equation of motion for the Goldstone modes on the branes is given by
\begin{eqnarray}
\partial_a 
\Big(\sqrt{-q} \Phi^2 
q^{ab}\big(\partial_b \sigma -eA_b \big) \Big)
=0\,,
\end{eqnarray}
which implies $\partial_{\theta}^2\sigma_{\pm}=0$.
The equation of motion for the axial mode $\sigma$ can be solved as
$\sigma_{\pm} =n_{\pm}\theta$.
From Eq. (\ref{vectorpotential}), (\ref{junction4}), (\ref{X}),
 one finds that the brane quantum numbers $n_{\pm}$ are given by
\begin{eqnarray}
&&n_{+}
=\frac{N}{2}
\frac{5-8\alpha^3+3\alpha^8}
     {4(1-\alpha^3)(1-\alpha^5)}\,,
\nonumber\\
&&n_{-}
=-\frac{N}{2}
\frac{5-8\alpha^{-3}+3\alpha^{-8}}
     {4(1-\alpha^{-3})(1-\alpha^{-5})}\,,
\label{quantum numbers}
\end{eqnarray}
where
we have also used Eq. (\ref{dirac}).
Thus, between quantum numbers, we have the following relations:
\begin{eqnarray}
n_{+}-n_{-}= N.
\end{eqnarray}
Note that the same relations as Eq. (\ref{quantum numbers})
are obtained in the case of the static brane.
Due to the redundancy of the junction condition, they
appear as a constraint between the model parameters.

\hspace{0.5cm}

\subsection{Einstein-Maxwell-dilaton model}

The dynamics of the radial scalar field on the brane is given by
\begin{eqnarray}
&&\xi \ddot{\Phi}
+\xi \dot{A}\Big(\frac{3}{A}+\frac{C'(A)}{C}\Big)
\dot{\Phi}
+\xi{}' \dot{\varphi} \dot{\Phi}
+V_{\Phi}
+\frac{\xi}{C^2}
 \Big(\partial_{\theta} \sigma
-eA_{\theta} \Big)^2\Phi
\nonumber \\
&&=0\,.
\end{eqnarray}
The equation of motion of $\Phi$ field is consistent with the energy momentum conservation law on the brane:
\begin{eqnarray}
 \dot{\Sigma}
+\frac{3\dot{A}}{A}
 \Big(\Sigma+p\Big)
+\frac{C' \dot{A}}{C}
\Big(\Sigma+p_{\theta}\Big)
=\Big[{\cal T}_{AB}u^A n^B\Big]\,.
\end{eqnarray}
where
\begin{eqnarray}
 \Big[{\cal T}_{AB}u^A n^B\Big]
&=&\dot{A}\Big\{e^{-\varphi}\Big(F_{A\theta}\Big)^2 C^{-2}
 +\Big(\varphi_{,A}\Big)^2\Big\} \Big[X\Big]
\nonumber\\
&=&-e\xi \dot{A}C^{-2}
 \Phi^2 
\big(\partial_{\theta}\sigma
-eA_{\theta}\big)
e^{-\varphi} F_{A\theta}
+\varphi_{,A}V_{,\varphi}\dot{A}
\nonumber\\
&+&\frac{1}{2}
 \xi'(\varphi) \varphi_{,A}\dot{A}
 \Big(-\dot{\Phi}^2
+ \Phi^2C^{-2} (\partial_{\theta}\varphi-eA_{\theta})\Big)\,,
\nonumber\\
\end{eqnarray}
and at the final step we used the Maxwell and dilaton junction conditions.

The equation of motion for the Goldstone modes on the branes is again given by
\begin{eqnarray}
\partial_a 
\Big\{\sqrt{-q}\xi(\varphi)  \Phi^2 
q^{ab}\big(\partial_b \sigma -eA_b \big) \Big\}
=0\,,
\end{eqnarray}
which implies $\partial_{\theta}^2\sigma=0$.
The solution is also given by
$\sigma=n_{\pm}\theta$.
The quantum numbers $n_{\pm}$ are found to be given by
\begin{eqnarray}
n_{\pm}=\pm\frac{N}{2}\,,
n_+ - n_-=N\,.
\end{eqnarray}




\end{document}